\documentclass[twoside]{class-definitions}
\usepackage[latin1]{inputenc}
\usepackage[dvips]{graphicx,epsfig,color}
\usepackage{wrapfig,rotating}
\usepackage{amssymb,amsmath,array}

\newcommand{\Wqq} {$W \rightarrow q\bar{q}$ }
\newcommand{\Zqq} {$Z \rightarrow q\bar{q}$ }
\newcommand{\Wev} {$W \rightarrow e \nu$ }
\newcommand{\Wuv} {$W \rightarrow \mu \nu$ }

\newcommand{\Zuu} {$Z \rightarrow \mu^+ \mu^-$ }
\newcommand{\dream} {{\sc dream }}
\newcommand{\C}    {\^{C}erenkov }   
\newcommand{\fem} {$f_{\sc em}$ }

\begin{document}
\title{The 4th Concept Detector \\  for the International Linear Collider} 
\author{Sung Keun Park$^{1,\footnote{Asian Contact for 4th Concept.}}$, Franco Grancagnolo$^2$, \\
  John Hauptman$^3$, Alexander Mikhailichenko$^4$, and Nural Akchurin$^5$
\thanks{This work has been supported by the US Department of Energy and by INFN, Italy.}
\vspace{.3cm}\\
1 - {\it Korea University, Department of Physics}, Anam-dong, Seoul 136-701, Korea \\
2 - {\it INFN and Dipartimento di Fisica}, via Lecce-Arnesano, 731000, Lecce, Italy \\
3 - {\it Iowa State University, Department of Physics and Astronomy},
Ames, IA  50011  USA \\
4- {\it Laboratory for Elementary Particle Physics, Cornell University}, Ithaca, NY 14853-5001 USA \\
5 - {\it Texas Tech University, Department of Physics}, Lubbock, TX 79409-1051 USA
} 

\maketitle

{\footnotesize
\centerline{Daniele Barbareschi, Emanuela Cavallo, Vito Di Benedetto, }
\centerline{ Corrado Gatto, Fedor Ignatov, Anna Mazzacane,
                 Giovanni Tassielli, Giuseppina Terracciano}  
\centerline{{\it INFN and Dipartimento di Fisica}, {\rm via Lecce-Arnesano, 73100,   Lecce, Italy}}

\centerline{Antonio Lamberto,  Gaetana Francesca Rappazzo, Ada Anania, {\it INFN, Messina, Italy}  }

\centerline{Corrado Gatto, {\it INFN, Naples,   {\rm Italy}} }

\centerline{Gianluca Introzzi, {\it INFN, Pavia,   {\rm Italy}} }

\centerline{Franco Bedeschi, Roberto Carosi, Marco Incagli, {\it INFN, Pisa,   {\rm Italy}} }

\centerline{Valter Bonvicini, Aldo Penzo, Irina Rashevskaya, Erik Vallazza, Gianluca Zampa}
\centerline{\it INFN, Trieste, Padriciano 99; I-34012 Padriciano, 
                      {\rm Trieste, Italy}}

\centerline{D. Cauz, C. Delpapa, G. Pauletta, M. Rossi, L. Santi}
\centerline{{\it  Univ. of Udine and INFN Ts. - G.C. Udine,  Vial delle Scienze}, Udine 33100. Italy}

\centerline{Sunghwan Ahn, Tae Jeong Kim, Kyong Sei Lee,}
\centerline{{\it Department of Physics, Korea University},
                         {\rm Seoul 136-701, Korea}}
                         
\centerline{Sorina Popescu, Laura Radulescu$^3$, {\it IFIN-HH, \rm Bucharest, Romania}}

\centerline{Sezen Sekmen, Efe Yazgan$^2$, Mehmet Zeyrek}
\centerline{{\it Physics Department, Middle East Technical University},
                    {\rm Ankara, Turkey}}

\centerline{S.I. Bondarenko, A.N. Omeliyanchuk, A.A. Shablo, N.S. Scherbakova,
     N.M.  Levchenko}
\centerline{ {\it  Institute for Low temperature Physics and Engineering},
                                {\rm Kharkov, Ukraine} }

\centerline{Muzaffer Atac, Marcel Demarteau, Ingrid Fang, Stephen R. Hahn, }
\centerline{Caroline Milstene, Erik Ramberg,  Robert Wands, Ryuji Yamada,  G.P. Yeh}
\centerline{\it Fermi National Accelerator Laboratory, 
                                   \rm Batavia, IL  60510  USA}

\centerline{Anatoli Frishman,  Jerry Lamsa, Sehwook Lee, Norio Nakagawa, German Valencia}
\centerline{\it Department of Physics and Astronomy, Iowa State University,
                      {\rm Ames, IA  50011 USA} }

\centerline{Heejong Kim, Sungwon Lee, Mario Spezziga$^3$, 
                         Igor Volobouev}  
\centerline{\it Department of Physics, Texas Tech University}
\centerline{\rm Lubbock, TX  79409-3783 USA}
}



\bigskip \bigskip \centerline{\bf Abstract} \bigskip

\begin{abstract}
The 4th Concept detector presently being designed for the International 
Linear Collider introduces several innovations in order
to achieve the necessary experimental goal of a detecter that is 2-to-10 
times better than the already excellent SLC and LEP detectors.  We introduce a
dual-readout calorimeter system, a cluster counting drift chamber, 
and a second solenoid to return the magnetic flux without iron.   We discuss
particle identification, momentum and energy resolutions, and the machine-detector
interface that together offer the possibility of a very high-performance detector for
$e^+e^-$physics up to $\sqrt{s} = 1$ TeV.  
\end{abstract}

\newpage
\section{Introduction}

The physics reach of  a new  high energy $e^+e^-$ linear  collider requires \cite{DCR} the 
detection of all known partons of the standard model ($e, \mu, \tau, uds, c, b, W, Z, \gamma, \nu$) 
 including the hadronic decays of the
gauge bosons, \Wqq and \Zqq and, by subtraction, the missing neutrinos in \Wev, \Wuv,
$\tau \rightarrow \ell \nu_{\ell} \nu_{\tau}$ and $\tau \rightarrow h \nu_{\tau}$  decays
so that kinematically over-constrained final states can be achieved.  The main
benchmark process is
\begin{displaymath}
 e^+e^- \rightarrow H^0 Z^0 \rightarrow (anything) + \mu^+\mu^-
\end{displaymath}
in which the two $\mu$s are measured in the tracking system and the Higgs is
seen in the missing mass distribution against the $\mu^+\mu^-$ system.  A
momentum resolution of $\sigma_p/p^2 \approx 4 \times 10^{-5}$ (GeV/c)$^{-1}$
is required for a desired  Higgs mass resolution of 150 MeV/c$^2$ in a 500 fb$^{-1}$
data sample.  There are
three main technologies under study to achieve this performance:
a 5-layer silicon strip tracker, a TPC with sophisticated high-precision end planes,
and a cluster-counting drift chamber \cite{Trk-Review}.

This same final state can be studied for \Zqq  decays which are 20 times more
plentiful than \Zuu decays but less distinct experimentally.  In addition, those 
processes that produce $W$ and $Z$ bosons either by production 
$(e^+e^- \rightarrow W^+W^-, Z^0Z^0, HHZ)$ or by decay $(H \rightarrow W^+W^-)$,
will reply critically on the direct mass resolution on \Zqq and \Wqq decays, and
these processes demand that the calorimeter energy
resolution be $\sigma_E/E \approx 30\%/\sqrt{E}$ with a constant term less
than 1\%.  There are two main technologies under study:  
a highly segmented calorimeter volume with approximately (1 cm)$^3$ channels
for the implementation of Particle Flow Analysis (PFA) algorithms, and dual-readout
optical calorimeters that measure both scintillation and Cerenkov light \cite{Cal-Review}.

The identification of $b$ and $c$ quark and $\tau$ lepton decays is critical
to good physics since these massive particles are a gateway to the decays of
more massive, and possibiy new, particles \cite{Vtx-Review}.   The measurement of 
their respective decay lengths places stringent conditions on the spatial
precision of a vertex chamber and how close it can be to the beam interaction
spot.  Typically, a spatial resolution of $\sigma \approx 5 ~\mu$m is desired,
with large solid angular coverage.   The inner radius is limited by the debris
from beamstrahlung that is only suppressed by the axial tracking magnetic
field; the uncharged debris cannot be suppressed.

The region beyond the calorimeter is reserved for, typically, a hadron absorber
or muon filter to supplement the absorbing calorimeter.  This absorber is always
iron to provide mechanical support for the detector elements inside and to
return the magnetic flux generated by a superconducting solenoid that establishes 
the uniform momentum tracking field.  There are two main types of muon systems
under study:  an iron absorber interspersed with tracking chambers to measure
the trajectories of penetrating tracks, and an iron-free design in which the magnetic
flux is returned by a second outer solenoid.

\section{The 4th Concept}

\begin{wrapfigure}{r}{0.6\columnwidth} 
\centerline{\includegraphics[width=0.55\columnwidth]{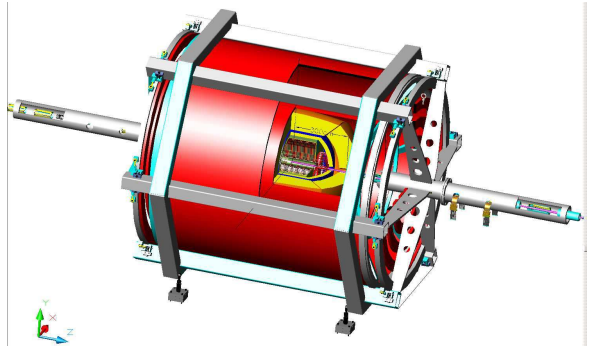}}
\caption{The 4th Concept detector.}\label{Fig:4th}
\end{wrapfigure}

We have introduced new ideas and instruments in order to achieve the resolution
requirements needed for ILC physics studies \cite{DOD}.  With the exception of the vertex
chamber, we have departed from the detectors at SLC and LEP and from the three
ILC concept detectors {\sc gld, ldc,} and {\sc sid} in all three major
detector systems:  the tracking, the calorimeter and the muon system.  The 4th detector
is displayed in Figure \ref{Fig:4th}, showing the beam transport through the final focus,
the dual solenoids (red), the crystal and fiber dual-readout calorimeters (yellow), the
tracking chamber, and the vertex chamber.   The muon tracking is in the annulus
between the solenoids.  A forward toroid for high precision forward tracking in under
current study in this figure.  Without iron, this detector is about 1/10 the mass of a 
conventional detector.

\subsection{Tracking by cluster counting in a low-mass He-based drift chamber}

\begin{wrapfigure}{r}{0.45\columnwidth} 
\centerline{\includegraphics[width=0.45\columnwidth]{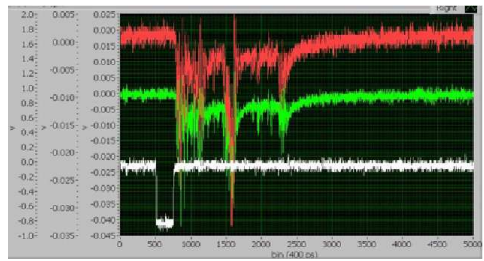}}
\caption{Ionization cluster data.}\label{Fig:clusters}
\end{wrapfigure}

The gaseous central tracker is under study as a cluster-counting drift chamber
modelled on the successful KLOE main tracking chamber  \cite{Trk-Review,pisa}.  This drift chamber 
(CluCou) maintains very low multiple scattering due to a He-based gas and aluminum 
wires in the tracking volume and utilizes carbon fiber end plates. Forward tracks 
(beyond $\cos \theta \approx 0.7$) which  penetrate the wire support frame and 
electronics pass through only about 15-20\% $X_{0}$ of material.   The low mass of 
the tracker directly improves momentum resolution in the multiple scattering dominated region 
below 10 GeV/c.  The He gas has a low drift velocity which allows a new cluster 
counting technique that clocks in individual ionization clusters on every 
wire, providing an estimated 50 micron spatial resolution per point, a $dE/dx$ resolution 
near 3\%, $z$-coordinate information on each track segment through an effective dip 
angle measurement,  and a layout made exclusively of super-layers with alternated opposite sign stereo angles.  The maximum drift time in each cell is less than the 300 ns beam crossing 
interval, so this chamber sees only one crossing per readout.  Data from a test of cluster counting
are shown in Figure \ref{Fig:clusters}.

The critical issues of occupancy and two-track resolution are being simulated for ILC 
events and expected machine and event backgrounds, and direct GHz cluster counting 
experiments are being performed.  This chamber has timing and pattern recognition
capabilities midway between the faster, higher precision silicon tracker and the slower, full imaging 
TPC, and is superior to both with respect to its low multiple scattering.  

The low-mass of the tracking medium, the multiplicity of point measurements, and the point
spatial precision allow this chamber to reach $\sigma_p/p^2 \approx 5 \times 10^{-5}$ (GeV/c)$^{-1}$
at high momenta, and to maintain good momentum resolution down to low momenta.

\subsection{Calorimetry by dual-readout of scintillation and \C light}

\begin{wrapfigure}{r}{0.55\columnwidth} 
\centerline{\includegraphics[width=0.50\columnwidth]{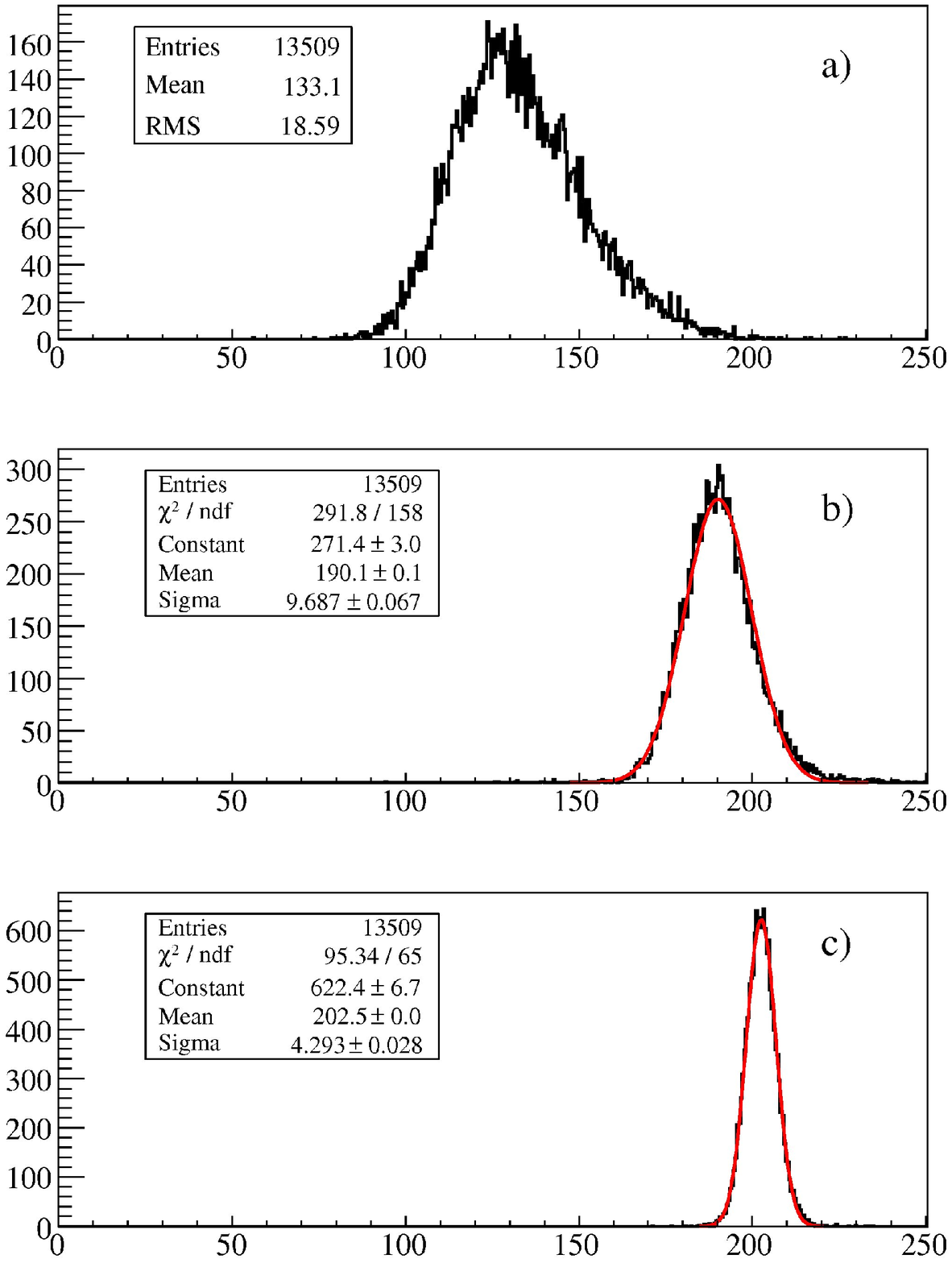}}
\caption{\dream energy resolutions.}\label{Fig:Eres-3}
\end{wrapfigure}

The calorimeter is a spatially fine-grained dual-readout fiber
sampling calorimeter augmented with the ability to measure the 
neutron content of a shower.  The dual fibers are sensitive to scintillation and 
Cerenkov radiation, for separation of the hadronic and electromagnetic components of 
hadronic showers \cite{dream-ha}, and since fluctuations in the {\sc em} fraction, i.e.,
fluctuations in $\pi^0$ production, are largely responsible for poor hadronic energy
resolution, the \dream module achieved a substantial improvement in hadronic
calorimetry.  The energy
resolution of the tested \dream calorimeter should be surpassed with finer spatial 
sampling, neutron detection for the measurement of fluctuations
in binding energy losses\cite{dream-n}, and a larger volume test module to reduce
leakage fluctuations.  The calorimeter modules
will have fibers up to their edges, and will be constructed for sub-millimeter
close packing, with signal extraction at the outer radius so that the
calorimeter system will approach full coverage without cracks.
A dual-readout crystal in front of the deep fiber module consists
 of a crystal calorimeter
with  readout of both scintillation and \C light\footnote{The \dream collaboration has
recently succeeded in dual-readout of a single BGO crystal and will publish these
results shortly.  Separation in the more difficult PWO crystal\cite{dream-PWO} 
has also been  accomplished .}. This provides 
better photoelectron statistics and therefore achieves better 
energy and spatial resolution for photons and electrons than is 
possible in the fiber calorimeter modules.  The dual
readout of these crystals is essential: over one-half of all hadrons
interact in the so-called {\sc em} section, depositing widely 
fluctuating fractions of {\sc em} and hadronic energy losses.

The fiber calorimeter shows promise of excellent energy resolution on
hadrons and jets, as seen in Figure \ref{Fig:Eres-3} for 200 GeV   $\pi^-$:
(a) the distribution of the scintillator ($S$) signal shows the raw resolution 
that a typical scintillating   sampling  calorimeter would achieve; (b) shows 
the leakage-dominated   energy distribution using only the scintillating $(S)$
and \C $(C)$ signals  for each event; and, 
   (c) shows the energy distribution with leakage fluctuations suppressed 
   using the known beam energy (=200 GeV)  to make a better estimate
   of the {\sc em} fraction each event.  The actual energy resolution of a fiber dual-readout
   calorimeter lies between Figures (b) and (c).

\begin{wrapfigure}{r}{0.6\columnwidth}
\centerline{\includegraphics[width=0.55\columnwidth]{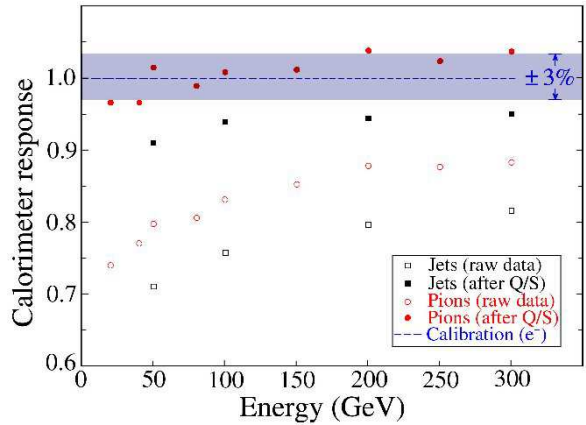}}
\caption{Measured response of the dual readout calorimeter for
  hadrons from 20 to 300 GeV.  The \dream module was calibrated
  only with 40 GeV electrons. }\label{Fig:linearity}
\end{wrapfigure}

Finally, and very importantly, the hadronic response of this 
dual-readout calorimeter is demonstrated to be linear in hadronic
energy from 20 to 300 GeV having been {\it calibrated only with 
40 GeV electrons}, Fig. \ref{Fig:linearity}. This is a critical advantage at the ILC where
calibration with 45 GeV electrons from $Z$ decay will suffice to maintain the energy
scale up to 10 times this energy for physics.

\subsection{Muon measurement without iron:  a dual solenoid configuration}

\begin{figure}[htb!]
 \epsfysize=6.0cm
 \centerline{\epsffile{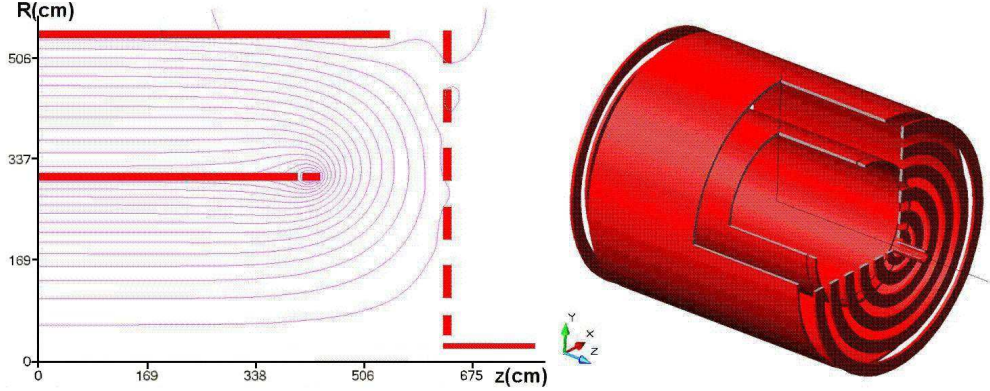}}
 \caption[The coils of the 4th concept and the field lines]{Drawings 
   showing the two solenoids and the ``wall of coils''
   that redirects the field out radially, and the resulting field lines in an
     $r-z$ view.  This field is uniform to 1\% at $3.5$ T in the tracking
     region, and also uniform and smooth at $-1.5$ T in the muon
     tracking annulus between the solenoids. }
 \label{fig:B+coils}
\end{figure}

The muon system utilizes a dual-solenoid magnetic field configuration 
in which the flux from the inner solenoid is returned
through the annulus between this inner solenoid and an 
outer solenoid oppositely driven with a smaller turn density \cite{Mik}.  
The magnetic field in the volume between 
the two solenoids will back-bend muons which have penetrated the calorimeter 
and allow, with the addition of tracking chambers, a second momentum measurement.
This will achieve high precision without the limitation of 
multiple scattering in $Fe$ that fundamentally limits
momentum resolution in conventional muon systems to 10\%.    
High spatial precision drift tubes with cluster counting
electronics are used to measure tracks in this volume  \cite{franco-mu-val}.  
The dual-solenoid field is terminated by a novel ``wall of coils''
that provides muon bending down to small angles 
($\cos \theta \approx 0.975$) and also
allows good control of the magnetic environment 
on and near the beam line. The design is illustrated in Fig~\ref{fig:B+coils}.

The path integral of the field in the annulus for  a muon from 
the origin is about 3 T$\cdot$m
over $0 < \cos \theta < 0.85$ and remains larger than 0.5 T$\cdot$m
out to $\cos \theta = 0.975$, allowing both good momentum resolution
and low-momentum acceptance over almost all of $4\pi$ \cite{franco-mu-val}.

For isolated tracks, the dual readout calorimeter independently provides a unique 
identification of muons relative to pions with a background track
rejection of $10^4$, or better, through its separate measurements
of ionization and radiative energy losses.

The detector's magnetic field is confined essentially to a cylinder with
negligible fringe fields, without the use of iron flux return. This scheme
offers flexibility in controlling the fields 
along the beam axis.  The twist compensation solenoid
just outside the wall of coils is shown in the above figure, along
with the beam line elements close to the IP.  This iron-free 
configuration \cite{Mik}
allows us to mount all beam line elements on a single support and
drastically reduce the effect of vibrations at the final focus (FF),
essentially because the beams will coherently move up and down together.
In addition, the FF elements can be brought close to the vertex chamber
for better control of the beam crossing.   

The open magnetic geometry of the 4th Concept also allows for future 
physics flexibility for asymmetric energy collisions, the installation
of specialized detectors outside the inner solenoid, and
magnetic flexibility for non-zero dispersion FF optics at the IP, 
adiabatic focussing at the IP, and monochromatization of the
collisions to achieve a minimum energy spread \cite{Mik}.
Finally, this flexibility and openness does not prevent additions
in later years  to the detector or to the beam line, and therefore no
physics  is precluded by this detector concept. 

\section{Particle Identification}

The capability to identify standard model  partons ($e, \mu, \tau, u/d/s,
c, b, t, W, Z, \gamma$) is equivalent to increased luminosity with larger
and less ambiguous data ensembles available for physics analysis.

\subsection{$\pi^0 \rightarrow \gamma \gamma$ separation and reconstruction}

The dual-readout crystals can be made small, about 1cm$\times$1cm or 2cm$\times$2cm,
and with reconstruction using shower shapes in the crystals we estimate that 
$\pi^0 \rightarrow \gamma \gamma$ can be reconstructed up to about $E_{\pi^0} \sim 20$ GeV,
which is high enough to reconstruct the important decay 
  $\tau \rightarrow \rho \nu \rightarrow \pi^{\pm} \pi^0 \nu \rightarrow \pi^{\pm} \gamma \gamma \nu$
to be used as a spin analyzer in the decays of higgs, etc.

\subsection{$e, \pi, K, p$ separation by $dE/dx$ at lower momenta}

The cluster counting central tracking chamber has the added benefit of an
excellent energy loss measurement without a Landau tail since clusters are
counted as opposed to energy losses summed.   We anticipate 3\% or
better resolution in $dE/dx$ as an analysis tool in, for example,  $b$ physics
where a large fraction  of charged tracks are below a few GeV/c.

\subsection{$\mu$ separation from $\pi^{\pm}$}

\begin{wrapfigure}{r}{0.6\columnwidth}
\centerline{\includegraphics[width=0.55\columnwidth]{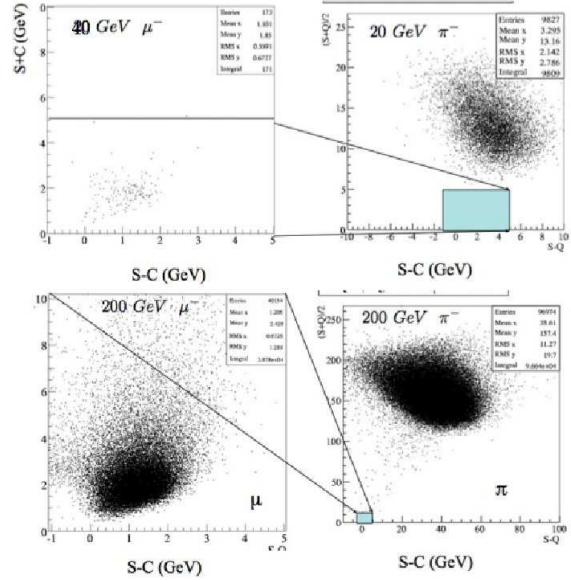}}
\caption{Dual-readout separation of $\mu-\pi^{\pm}$.}\label{Fig:mu-pi}
\end{wrapfigure}

 In the 4th
concept, we achieve excellent $\mu-\pi^{\pm}$ separation in the dual
readout calorimeter and additional separation using energy balance
from the tracker through the calorimeter into the muon spectrometer.
A non-radiating $\mu$ has a zero \C signal in the fiber calorimeter since
the \C angle is larger than the capture cone angle of the fiber.  The 
scintillating fibers measure $dE/dx$ of the throughgoing $\mu$ \cite{dream-mu}.
Any radiation by the $\mu$ within the body of the calorimeter is sampled
equally by the scintillating ($S$) and \C ($C$) fibers \cite{dream-e}, and therefore
$S-C \approx dE/dx$ independent of the amount of radiation.  The distributions of $(S-C)$
 {\it vs.} $(S+C)/2$ for 20 GeV and 200 GeV $\pi^-$ and $\mu^-$ are shown in Fig.
 \ref{Fig:mu-pi} in which it is evident that for an isolated track the $\pi^{\pm}$ 
 rejection against $\mu$ is about $10^4$ at 20 GeV and $10^5$ at 200 GeV.
 A further factor of 50 is obtained from the iron-free dual solenoid in which 
 the precisely measured $\mu$ momentum can be matched with the momentum
 in the central tracker and the radiated energy in the calorimeter. 
  Of course, we
 expect that other effects, such as tracking inefficiencies, will limit this level of
 rejection before these beam-measured rejections are achieved in practice.

 \subsection{$e$ separation from $\pi^{\pm}$ and jets $(j)$}

\begin{figure}[hbtp]
  \vspace{9pt}

  \centerline{\hbox{ \hspace{0.0in} 
    \epsfxsize=2.5in
    \epsffile{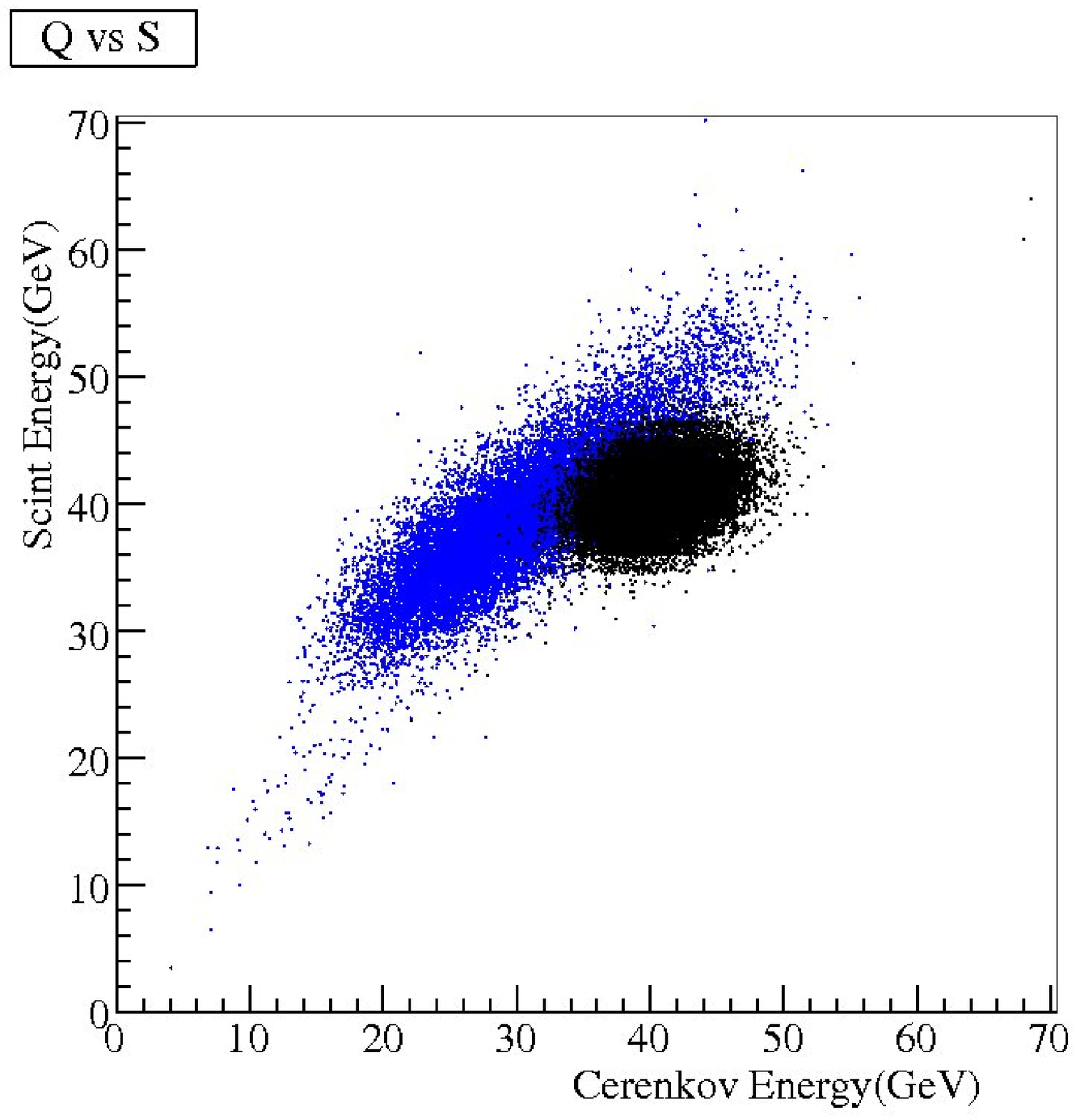}
    \hspace{0.25in}
    \epsfxsize=2.5in
    \epsffile{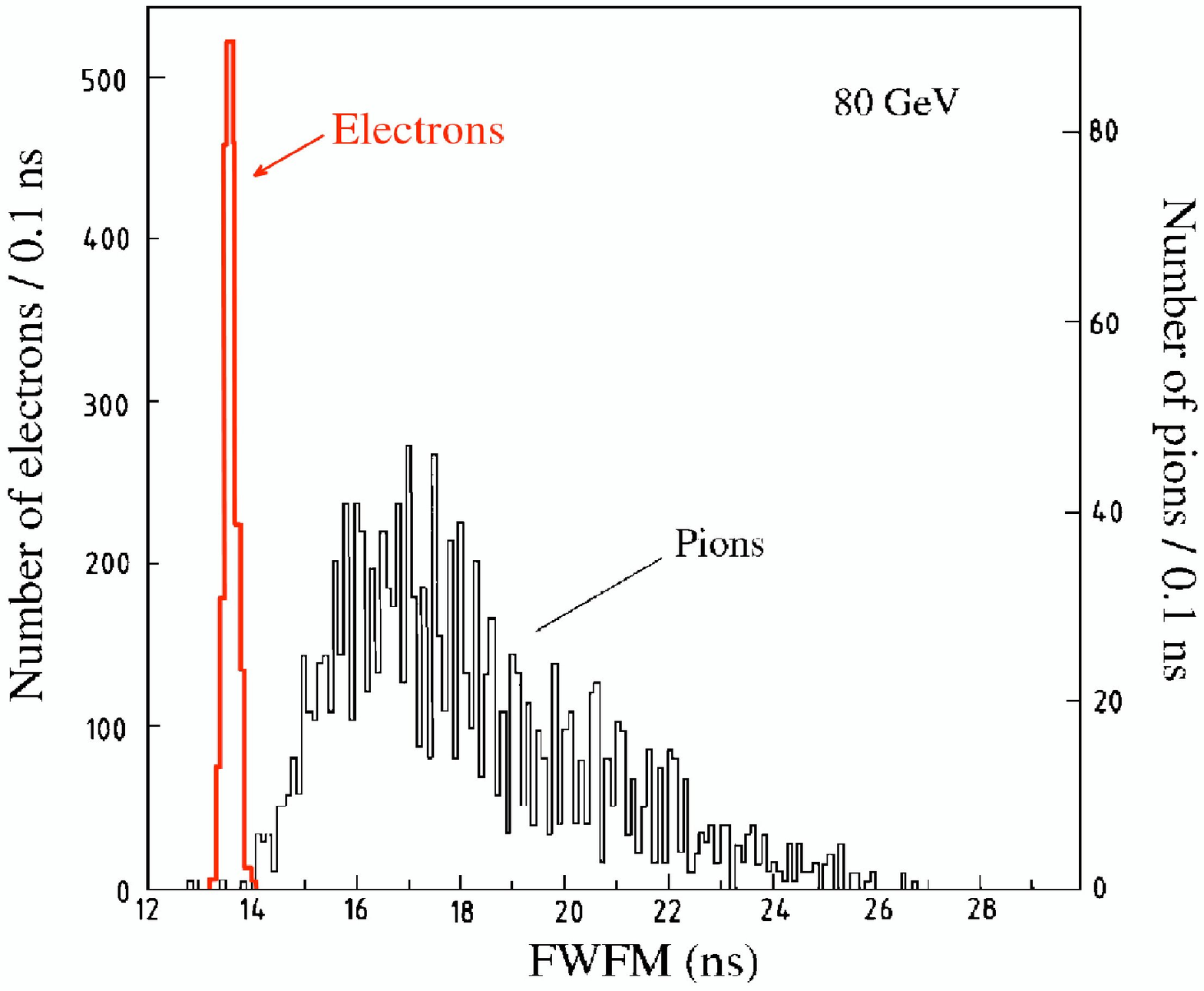}
    }
  }

  \vspace{9pt}
  \hbox{\hspace{1.35in} (a) \hspace{2.10in} (b)} 
  \vspace{9pt}

  \caption{(a) $C ~vs.~ S$ for 40 GeV $e^-$ and 50 GeV $\pi^-$; (b) $e- \pi$
  separation by time history.}
  \label{fig:e-pi}

\end{figure}


   The scintillation {\it vs.} the \C response of 40 GeV $e$ and 50 GeV $\pi^{\pm}$
   is displayed in Fig. \ref{fig:e-pi} in the \dream test module \cite{dream-ha,dream-e},
   showing the equal responses for $e$ and the \fem fluctuations for $\pi^{\pm}$, and the 
   obvious substantial rejection of $\pi^{\pm}$ against $e$.  Just as the overall shower
   $C ~vs.~ S$ response fluctuates, so do the individual channels of the $\pi^{\pm}$
   generated showers.  The statistic
 \begin{displaymath}
 \sigma_{Q-S} = \frac{1}{N}  \sum_{i=1}^{N} (Q_i - S_i)^2
 \end{displaymath}
 is a measure of these channel-to-channel fluctuations.  
 For 100 GeV $e$ this $\sigma^e \approx 0.2$ GeV$^2$, and 
 for $\pi^{\pm}$ $\sigma^{\pi} \approx 10$ GeV$^2$, yielding a rejection of
 $\pi^{\pm}$ against $e$ of about 50.
  
   The time history of the 
   scintillating signal contains independent information, in particular, the neutrons
   generated in the $\pi^{\pm}$ cascade travel slower ($v \sim 0.05 c$) and fill a
   larger volume, and therefore the elapsed time of the scintillating signal is longer
   for a $\pi^{\pm}$ and an $e$.  One statistic is the ``full width at one-fifth maximum''
   shown in Fig. \ref{fig:e-pi}(b) for 80 GeV $e$ and $\pi^{\pm}$ in the
   {\sc spacal} calorimeter \cite{spacal}.  These statistics will work just as well
   for $j$ separation from $e$  when the jet  has a single forward high energy
   $\pi^{\pm}$.
   
     The exploitation of these measurements in 
   a collider experiment will depend on many factors, such as channel size, the 
character of event ensembles, and the fidelity of the measurements themselves.  
The goal of the 4th Concept is to package these capabilities into   a comprehensive
detector.
  

\subsection{Time-of-flght with the dual-readout crystal and fiber calorimeters}

The time history readout of the scintillating fibers will serve several purposes,
{\it viz.}, $e-\pi$ separation (Fig. \ref{fig:e-pi}(b)), neutron measurement for the
suppression of fluctuations in  binding energy losses, and as a real-time
nanosecond monitor of all activity in the volume between the 337-ns bunch
crossings, including 'flyers' and beam burps of any kind.

In addition, the $ns$ and even sub-$ns$ resolution on the time of arrival of a
shower can be used in the time-of-flight of heavy objects such as
supersymmetric or technicolor particles that move slowly ($v \sim 0.1c$) 
into the tracking 
volume and decay to light particles ($e, \mu, \tau, j$, etc.).  Such objects 
can be easily reconstructed in 4th. 

\section{Summary}

The 4th Concept detector contains many new ideas in high energy physics
instrumentation that are aimed at a comprehensive detector for 1 TeV $e^+e^-$
physics at the International Linear Collider.  All of the data shown in this paper
and all of the performance specifications have already been tested in either
beam tests, prototypes, or existing detectors.  The difficult problems of 
incorporating these small successful instruments into a large detector 
while maintaining the scientific strengths of each are good work  in
the near future.



\begin{footnotesize}

\end{footnotesize}


\end{document}